

\font\twelverm=cmr10 scaled 1200    \font\twelvei=cmmi10 scaled 1200
\font\twelvesy=cmsy10 scaled 1200   \font\twelveex=cmex10 scaled 1200
\font\twelvebf=cmbx10 scaled 1200   \font\twelvesl=cmsl10 scaled 1200
\font\twelvett=cmtt10 scaled 1200   \font\twelveit=cmti10 scaled 1200
\skewchar\twelvei='177   \skewchar\twelvesy='60
\def\twelvepoint{\normalbaselineskip=12.4pt
  \abovedisplayskip 12.4pt plus 3pt minus 9pt
  \belowdisplayskip 12.4pt plus 3pt minus 9pt
  \abovedisplayshortskip 0pt plus 3pt
  \belowdisplayshortskip 7.2pt plus 3pt minus 4pt
  \smallskipamount=3.6pt plus1.2pt minus1.2pt
  \medskipamount=7.2pt plus2.4pt minus2.4pt
  \bigskipamount=14.4pt plus4.8pt minus4.8pt
  \def\rm{\fam0\twelverm}          \def\it{\fam\itfam\twelveit}%
  \def\sl{\fam\slfam\twelvesl}     \def\bf{\fam\bffam\twelvebf}%
  \def\mit{\fam 1}                 \def\cal{\fam 2}%
  \def\tt{\twelvett}
  \textfont0=\twelverm   \scriptfont0=\tenrm   \scriptscriptfont0=\sevenrm
  \textfont1=\twelvei    \scriptfont1=\teni    \scriptscriptfont1=\seveni
  \textfont2=\twelvesy   \scriptfont2=\tensy   \scriptscriptfont2=\sevensy
  \textfont3=\twelveex   \scriptfont3=\twelveex  \scriptscriptfont3=\twelveex
  \textfont\itfam=\twelveit
  \textfont\slfam=\twelvesl
  \textfont\bffam=\twelvebf \scriptfont\bffam=\tenbf
  \scriptscriptfont\bffam=\sevenbf
  \normalbaselines\rm}

\def\beginlinemode{\endmode
  \begingroup\parskip=0pt \obeylines\def\\{\par}\def\endmode{\par\endgroup}}
\def\beginparmode{\endmode
  \begingroup \def\endmode{\par\endgroup}}
\let\endmode=\par
{\obeylines\gdef\
{}}
\def\singlespace{\baselineskip=\normalbaselineskip}
\def\oneandthreefifthsspace{\baselineskip=\normalbaselineskip
  \multiply\baselineskip by 8 \divide\baselineskip by 5}

\def\oneandahalfspace{\baselineskip=\normalbaselineskip
  \multiply\baselineskip by 3 \divide\baselineskip by 2}
\def\doublespace{\baselineskip=\normalbaselineskip \multiply\baselineskip by 2}
\newcount\firstpageno
\firstpageno=2
\footline=
{\ifnum\pageno<\firstpageno{\hfil}\else{\hfil\twelverm\folio\hfil}\fi}
\let\rawfootnote=\footnote              
\def\footnote#1#2{{\rm\singlespace\parindent=0pt\rawfootnote{#1}{#2}}}
\def\raggedcenter{\leftskip=2em plus 12em \rightskip=\leftskip
  \parindent=0pt \parfillskip=0pt \spaceskip=.3333em \xspaceskip=.5em
  \pretolerance=9999 \tolerance=9999
  \hyphenpenalty=9999 \exhyphenpenalty=9999 }
\parskip=\medskipamount
\twelvepoint            
\overfullrule=0pt       
\def\preprintno#1{
 \rightline{\rm #1}}    
\def\author                     
  {\vskip 3pt plus 0.2fill \beginlinemode
   \singlespace \raggedcenter \twelvesc}
\def\affil                      
  {\vskip 3pt plus 0.1fill \beginlinemode
   \oneandahalfspace \raggedcenter \sl}
\def\abstract                   
  {\vskip 3pt plus 0.3fill \beginparmode
   \doublespace \narrower \noindent ABSTRACT: }
\def\endtitlepage               
  {\endpage                     
   \body}
\def\body                       
  {\beginparmode}               

\def\subhead#1{                 
  \vskip 0.25truein             
  {\raggedcenter #1 \par}
   \nobreak\vskip 0.1truein\nobreak}
\def\refto#1{$|{#1}$}           
\def\references                 
  {\subhead{References}         
   \beginparmode
   \frenchspacing \parindent=0pt \leftskip=1truecm
   \parskip=8pt plus 3pt \everypar{\hangindent=\parindent}}
\gdef\refis#1{\indent\hbox to 0pt{\hss#1.~}}    
\gdef\journal#1, #2, #3, 1#4#5#6{               
    {\sl #1~}{\bf #2}, #3, (1#4#5#6)}           
\def\refstylenp{                
  \gdef\refto##1{ [##1]}                                
  \gdef\refis##1{\indent\hbox to 0pt{\hss##1)~}}        
  \gdef\journal##1, ##2, ##3, ##4 {                     
     {\sl ##1~}{\bf ##2~}(##3) ##4 }}
\def\refstyleprnp{              
  \gdef\refto##1{ [##1]}                                
  \gdef\refis##1{\indent\hbox to 0pt{\hss##1)~}}        
  \gdef\journal##1, ##2, ##3, 1##4##5##6{               
    {\sl ##1~}{\bf ##2~}(1##4##5##6) ##3}}
\def\pr{\journal Phys. Rev., }

\def\prl{\journal Phys. Rev. Lett., }
\def\prpts{\journal Phys. Rep., }
\def\np{\journal Nucl. Phys., }

\def\pl{\journal Phys. Lett., }

\def\endreferences{\body}
\def\endpage                    
  {\vfill\eject}
\def\endpaper                   
  {\endmode\vfill\supereject}
\def\endit
  {\endpaper\end}
\def\ref#1{Ref. #1}                     
\def\Ref#1{Ref. #1}                     

\def\m@th{\mathsurround=0pt }
\font\twelvesc=cmcsc10 scaled 1200
\def\cite#1{{#1}}
\def\(#1){(\call{#1})}
\def\call#1{{#1}}
\def\taghead#1{}
\def\leaderfill{\leaders\hbox to 1em{\hss.\hss}\hfill}
\def\twiddle{\lower.9ex\rlap{$\kern-.1em\scriptstyle\sim$}}
\def\bigtwiddle{\lower1.ex\rlap{$\sim$}}
\def\gtwid{\mathrel{\raise.3ex\hbox{$>$\kern-.75em\lower1ex\hbox{$\sim$}}}}
\def\ltwid{\mathrel{\raise.3ex\hbox{$<$\kern-.75em\lower1ex\hbox{$\sim$}}}}
\def\square{\kern1pt\vbox{\hrule height 1.2pt\hbox{\vrule width 1.2pt\hskip 3pt
   \vbox{\vskip 6pt}\hskip 3pt\vrule width 0.6pt}\hrule height 0.6pt}\kern1pt}
\catcode`@=11
\newcount\tagnumber\tagnumber=0

\immediate\newwrite\eqnfile
\newif\if@qnfile\@qnfilefalse
\def\write@qn#1{}
\def\writenew@qn#1{}
\def\w@rnwrite#1{\write@qn{#1}\message{#1}}
\def\@rrwrite#1{\write@qn{#1}\errmessage{#1}}

\def\taghead#1{\gdef\t@ghead{#1}\global\tagnumber=0}
\def\t@ghead{}

\expandafter\def\csname @qnnum-3\endcsname
  {{\t@ghead\advance\tagnumber by -3\relax\number\tagnumber}}
\expandafter\def\csname @qnnum-2\endcsname
  {{\t@ghead\advance\tagnumber by -2\relax\number\tagnumber}}
\expandafter\def\csname @qnnum-1\endcsname
  {{\t@ghead\advance\tagnumber by -1\relax\number\tagnumber}}
\expandafter\def\csname @qnnum0\endcsname
  {\t@ghead\number\tagnumber}
\expandafter\def\csname @qnnum+1\endcsname
  {{\t@ghead\advance\tagnumber by 1\relax\number\tagnumber}}
\expandafter\def\csname @qnnum+2\endcsname
  {{\t@ghead\advance\tagnumber by 2\relax\number\tagnumber}}
\expandafter\def\csname @qnnum+3\endcsname
  {{\t@ghead\advance\tagnumber by 3\relax\number\tagnumber}}

\def\equationfile{%
  \@qnfiletrue\immediate\openout\eqnfile=\jobname.eqn%
  \def\write@qn##1{\if@qnfile\immediate\write\eqnfile{##1}\fi}
  \def\writenew@qn##1{\if@qnfile\immediate\write\eqnfile
    {\noexpand\tag{##1} = (\t@ghead\number\tagnumber)}\fi}
}

\def\callall#1{\xdef#1##1{#1{\noexpand\call{##1}}}}
\def\call#1{\each@rg\callr@nge{#1}}

\def\each@rg#1#2{{\let\thecsname=#1\expandafter\first@rg#2,\end,}}
\def\first@rg#1,{\thecsname{#1}\apply@rg}
\def\apply@rg#1,{\ifx\end#1\let\next=\relax%
\else,\thecsname{#1}\let\next=\apply@rg\fi\next}

\def\callr@nge#1{\calldor@nge#1-\end-}
\def\callr@ngeat#1\end-{#1}
\def\calldor@nge#1-#2-{\ifx\end#2\@qneatspace#1 %
  \else\calll@@p{#1}{#2}\callr@ngeat\fi}
\def\calll@@p#1#2{\ifnum#1>#2{\@rrwrite{Equation range #1-#2\space is bad.}
\errhelp{If you call a series of equations by the notation M-N, then M and
N must be integers, and N must be greater than or equal to M.}}\else%
 {\count0=#1\count1=#2\advance\count1
by1\relax\expandafter\@qncall\the\count0,%
  \loop\advance\count0 by1\relax%
    \ifnum\count0<\count1,\expandafter\@qncall\the\count0,%
  \repeat}\fi}

\def\@qneatspace#1#2 {\@qncall#1#2,}
\def\@qncall#1,{\ifunc@lled{#1}{\def\next{#1}\ifx\next\empty\else
  \w@rnwrite{Equation number \noexpand\(>>#1<<) has not been defined yet.}
  >>#1<<\fi}\else\csname @qnnum#1\endcsname\fi}

\let\eqnono=\eqno
\def\eqno(#1){\tag#1}
\def\tag#1$${\eqnono(\displayt@g#1 )$$}

\def\aligntag#1\endaligntag
  $${\gdef\tag##1\\{&(##1 )\cr}\eqalignno{#1\\}$$
  \gdef\tag##1$${\eqnono(\displayt@g##1 )$$}}

\def\eqalignno#1{\displ@y \tabskip\centering
  \halign to\displaywidth{\hfil$\displaystyle{##}$\tabskip\z@skip
    &$\displaystyle{{}##}$\hfil\tabskip\centering
    &\llap{$\displayt@gpar##$}\tabskip\z@skip\crcr
    #1\crcr}}

\def\displayt@gpar(#1){(\displayt@g#1 )}

\def\displayt@g#1 {\rm\ifunc@lled{#1}\global\advance\tagnumber by1
        {\def\next{#1}\ifx\next\empty\else\expandafter
        \xdef\csname @qnnum#1\endcsname{\t@ghead\number\tagnumber}\fi}%
  \writenew@qn{#1}\t@ghead\number\tagnumber\else
        {\edef\next{\t@ghead\number\tagnumber}%
        \expandafter\ifx\csname @qnnum#1\endcsname\next\else
        \w@rnwrite{Equation \noexpand\tag{#1} is a duplicate number.}\fi}%
  \csname @qnnum#1\endcsname\fi}

\def\ifunc@lled#1{\expandafter\ifx\csname @qnnum#1\endcsname\relax}

\let\@qnend=\end\gdef\end{\if@qnfile
\immediate\write16{Equation numbers written on []\jobname.EQN.}\fi\@qnend}

\catcode`@=12
\refstyleprnp
\catcode`@=11
\newcount\r@fcount \r@fcount=0
\def\refreset{\global\r@fcount=0}
\newcount\r@fcurr
\immediate\newwrite\reffile
\newif\ifr@ffile\r@ffilefalse
\def\w@rnwrite#1{\ifr@ffile\immediate\write\reffile{#1}\fi\message{#1}}

\def\writer@f#1>>{}
\def\referencefile{
  \r@ffiletrue\immediate\openout\reffile=\jobname.ref%
  \def\writer@f##1>>{\ifr@ffile\immediate\write\reffile%
    {\noexpand\refis{##1} = \csname r@fnum##1\endcsname = %
     \expandafter\expandafter\expandafter\strip@t\expandafter%
     \meaning\csname r@ftext\csname r@fnum##1\endcsname\endcsname}\fi}%
  \def\strip@t##1>>{}}

\def\citeall#1{\xdef#1##1{#1{\noexpand\cite{##1}}}}
\def\cite#1{\each@rg\citer@nge{#1}}     

\def\each@rg#1#2{{\let\thecsname=#1\expandafter\first@rg#2,\end,}}
\def\first@rg#1,{\thecsname{#1}\apply@rg}       
\def\apply@rg#1,{\ifx\end#1\let\next=\relax
\else,\thecsname{#1}\let\next=\apply@rg\fi\next}

\def\citer@nge#1{\citedor@nge#1-\end-}  
\def\citer@ngeat#1\end-{#1}
\def\citedor@nge#1-#2-{\ifx\end#2\r@featspace#1 
  \else\citel@@p{#1}{#2}\citer@ngeat\fi}        
\def\citel@@p#1#2{\ifnum#1>#2{\errmessage{Reference range #1-#2\space is bad.}%
    \errhelp{If you cite a series of references by the notation M-N, then M and
    N must be integers, and N must be greater than or equal to M.}}\else%
 {\count0=#1\count1=#2\advance\count1
by1\relax\expandafter\r@fcite\the\count0,%
  \loop\advance\count0 by1\relax
    \ifnum\count0<\count1,\expandafter\r@fcite\the\count0,%
  \repeat}\fi}

\def\r@featspace#1#2 {\r@fcite#1#2,}    
\def\r@fcite#1,{\ifuncit@d{#1}
    \newr@f{#1}%
    \expandafter\gdef\csname r@ftext\number\r@fcount\endcsname%
                     {\message{Reference #1 to be supplied.}%
                      \writer@f#1>>#1 to be supplied.\par}%
 \fi%
 \csname r@fnum#1\endcsname}
\def\ifuncit@d#1{\expandafter\ifx\csname r@fnum#1\endcsname\relax}%
\def\newr@f#1{\global\advance\r@fcount by1%
    \expandafter\xdef\csname r@fnum#1\endcsname{\number\r@fcount}}

\let\r@fis=\refis                       
\def\refis#1#2#3\par{\ifuncit@d{#1}
   \newr@f{#1}%
   \w@rnwrite{Reference #1=\number\r@fcount\space is not cited up to now.}\fi%
  \expandafter\gdef\csname r@ftext\csname r@fnum#1\endcsname\endcsname%
  {\writer@f#1>>#2#3\par}}

\def\ignoreuncited{
   \def\refis##1##2##3\par{\ifuncit@d{##1}%
   \else\expandafter\gdef\csname r@ftext\csname r@fnum##1\endcsname\endcsname%
     {\writer@f##1>>##2##3\par}\fi}}

\def\r@ferr{\endreferences\errmessage{I was expecting to see
\noexpand\endreferences before now;  I have inserted it here.}}
\let\r@ferences=\references
\def\references{\r@ferences\def\endmode{\r@ferr\par\endgroup}}

\let\endr@ferences=\endreferences
\def\endreferences{\r@fcurr=0
  {\loop\ifnum\r@fcurr<\r@fcount
    \advance\r@fcurr by 1\relax\expandafter\r@fis\expandafter{\number\r@fcurr}%
    \csname r@ftext\number\r@fcurr\endcsname%
  \repeat}\gdef\r@ferr{}\global\r@fcount=0\endr@ferences}

\let\r@fend=\endpaper\gdef\endpaper{\ifr@ffile
\immediate\write16{Cross References written on []\jobname.REF.}\fi\r@fend}

\catcode`@=12

\citeall\refto          
\citeall\ref            %
\citeall\Ref            %

\referencefile

\def\sI{\scriptscriptstyle I}
\def\sJ{\scriptscriptstyle J}
\def\sK{\scriptscriptstyle K}
\def\sL{\scriptscriptstyle L}
\def\sM{\scriptscriptstyle M}
\def\sN{\scriptscriptstyle N}
\def\frac#1/#2{#1 / #2}

\def\oneandthreefifthsspace{\baselineskip=\normalbaselineskip
  \multiply\baselineskip by 8 \divide\baselineskip by 5}

\font\titlefont=cmr10 scaled\magstep3
\def\bigtitle                      
  {\null\vskip 3pt plus 0.2fill
   \beginlinemode \doublespace \raggedcenter \titlefont}


\oneandahalfspace
\preprintno{IASSNS-HEP-95/85}
\preprintno{UM-TH-95-25}
\preprintno{hep-ph/9510370}
\bigtitle{Flat directions in the scalar potential of
          the supersymmetric standard model}
\bigskip

\author Tony Gherghetta$^1$, Chris Kolda$^2$ and Stephen P.~Martin$^1$

\affil{$^1$Randall Physics Laboratory, University of Michigan, %
Ann Arbor MI 48109}
\affil{$^2$School of Natural Sciences, Institute for Advanced Study, %
Princeton NJ 08540}

\body

\abstract \oneandthreefifthsspace
The scalar potential of the Minimal Supersymmetric Standard Model (MSSM) is
nearly flat along many directions in field space. We provide a catalog of the
flat directions of the renormalizable and supersymmetry-preserving part of the
scalar potential of the MSSM, using the correspondence between flat directions
and gauge-invariant polynomials of chiral superfields. We then study how these
flat directions are lifted by non-renormalizable terms in the superpotential,
with special attention given to the subtleties associated with the family
index structure. Several flat directions are lifted only by
supersymmetry-breaking effects and by supersymmetric terms in the scalar
potential of surprisingly high dimensionality.

\endtitlepage
\oneandahalfspace

\subhead{1. Introduction}
\taghead{1.}

Supersymmetric gauge theories often possess a remarkable vacuum
degeneracy at the classical level. The renormalizable
scalar potential in supersymmetry
is a sum of squares of $F$-terms and $D$-terms, and so may vanish
identically along certain ``flat directions'' in field space. The space
of all such flat directions is called the moduli space, and the
massless chiral superfields whose expectation values parameterize
the flat directions are known as moduli. The properties of the
space of flat directions of a supersymmetric model are crucial
considerations for cosmology and whenever the behavior of the
theory at large field strengths is an issue.

In realistic models such as the Minimal Supersymmetric Standard Model
[\cite{reviews}]
(MSSM), the ``flat'' directions are only approximately flat; the
vacuum degeneracy of the scalar potential is lifted by soft
supersymmetry-breaking terms, and by non-renormalizable terms
in the superpotential. The soft terms contribute terms to the
scalar potential which are schematically of the form
$$
V_{\rm soft}  = m^2 |\phi |^2
\eqno(soft)
$$
where $\phi$ represents the scalar component of the moduli fields.
Now, if supersymmetry is to provide a successful explanation for the
hierarchy problem associated with the mass of the Higgs scalar boson,
$m$ must be of the order of the electroweak scale. The terms
in \(soft)  can never be forbidden by any symmetry (unlike soft terms
of the form $\phi^2$ and $\phi^3$), and so we expect that all flat
directions will be lifted weakly in this way.

The question of which non-renormalizable terms in the superpotential
also lift a given flat direction is more complicated. It is useful to
think of the non-renormalizable superpotential as an expansion in
inverse powers of some large mass scale $M$ (presumably the
Planck scale or some other physical cutoff); schematically
$$
W = W_{\rm renorm} + \sum_{n>3} {\lambda \over M^{n-3} } \Phi^n
\> .
\eqno(power)
$$
Each flat direction may be labeled by an order parameter modulus $\phi$
which can take on values with $|\phi| < M$. Therefore
it is sufficient to consider separately the contributions to the
superpotential first from renormalizable terms $W_{\rm renorm}$ and
then for each value of $n>3$ in turn. Renormalizable flat directions
are those for which all $F$-terms following from $W_{\rm renorm}$
and all $D$-terms
vanish.
Of these renormalizable flat directions, some are lifted when
$F$-terms from the $n=4$ superpotential are included;
some may survive until $n=5$ superpotential terms are included, etc.
The lowest order contribution to the scalar potential for a modulus
$\phi$ corresponding to a flat direction lifted at level $n$ is
$$
V \supset |F|^2 \sim {|\lambda |^2\over M^{2 n - 6}} | \phi |^{2 n -2}
\eqno(hard)
$$
so that (nearly) flat directions which survive to higher $n$ are
at least formally
``flatter''. We may therefore describe the relative flatness of
a renormalizable flat direction at large field strength by specifying
the level $n$ at which it is first lifted by non-renormalizable terms
in the superpotential.
The non-renormalizable
terms \(hard) dominate over the soft supersymmetry breaking terms
\(soft) in lifting
flat directions for field strengths
$$
\left ({m M^{n-3}\over |\lambda|}\right )^{1/(n-2)}< |\phi | < M \> .
\eqno(domrange)
$$
The scalar potential should also include
supersymmetry-breaking terms of the form
$$
V \supset {m \over M^{s-3}} \phi^s + c.c.
\eqno(softholo)
$$
where again $m$ must be of the order of the electroweak scale if
supersymmetry is to solve the hierarchy problem.
A phase rotation on $\phi$ can always make terms of this type negative.
However, these terms cannot dominate over both
\(soft) and \(hard) for renormalizable flat directions except when
$s<n$ (and even then only
for a limited range of $|\phi |$). In the MSSM, $s<n$ does not occur
for any of the renormalizable
flat directions assuming ``generic'' values for all couplings,
as we shall see.
In an inflationary epoch, the $m$ appearing in
\(soft) and \(softholo) may be identified [\cite{DRT}] with the (much larger)
Hubble expansion parameter. In most of
our remaining discussion, we will concentrate
on the moduli space of the supersymmetric part of the scalar potential.
\phantom{\cite{BDFS}\cite{ADS}\cite{LT}\cite{taylor}
\cite{Seiberg}\cite{IS}\cite{PR}
\cite{ILS}\cite{GP}}

Consider a model with $N$ chiral superfields $X_I$ transforming
under the gauge group $G$ as a (in general reducible) representation
in which the generators of the Lie gauge algebra are matrices $T^A$.
In principle, one could attempt to find all flat directions  in the scalar
potential $V = (g_A^2/ 2) D^A D^A + \sum |F_{X_I}|^2$ by solving the
simultaneous nonlinear equations
$$
\eqalignno{
D^A & \equiv X^\dagger T^A X = 0, &(Dflat) \cr
F_{X_I} & \equiv {\partial W /\partial X_I }= 0 &(Fflat)
\cr}
$$
for the scalar components $X_I$.
However, a more useful and elegant way of characterizing the
moduli space relies on the correspondence [\cite{BDFS},\cite{ADS}] 
between flat directions
and gauge-invariant, holomorphic polynomials of the chiral superfields
$X_I$. In particular, the moduli space of $D$-flat directions is
parameterized by a finite set of gauge-invariant monomials
of the chiral superfields
which obey a finite set of redundancy relations. The smaller moduli space
of $D$- and $F$-flat
directions is parameterized by the same basis of monomials
and redundancy constraints subject to additional constraints
(some linear and some non-linear) following from $F_{X_I} = 0$
(see [\cite{LT}] and references therein).
This result is the foundation for most of our discussion in this
paper. Intuitively, the order
parameter which describes motion along a given flat
direction is the scalar component of the corresponding
gauge invariant polynomial of chiral superfields. Examples of this
correspondence between flat directions and the gauge invariant polynomials
abound in the literature (see for example [\cite{DRT}-\cite{GP}]).

In the absence of exact global symmetries which would make the
non-renormalizable superpotential non-generic, one expects
that all flat directions will be lifted at some finite $n$. The
number of $F$-constraints \(Fflat) is formally equal to the number of degrees
of freedom in the chiral superfields $X_I$. In addition, one has dim($G$)
real gauge constraints (although not all of these
constraints are always independent),
and generally even more degrees of freedom
can be absorbed in gauge-fixing. Therefore the system ought to be
overconstrained, typically leaving only the trivial solution
$X_I=0$ as the minimum of the supersymmetric part of the potential.
Of course, this counting presupposes that the $F=0$ constraints are
all non-trivial and independent, an assumption which clearly does not
hold at the renormalizable level in many theories.
Furthermore, the requirement of gauge invariance (and, in realistic
models, matter parity or equivalently R-parity invariance)
of the superpotential severely limits the non-renormalizable
superpotential terms for smaller values of $n$, and so
may not allow flat directions
to be lifted except for some (perhaps large) $n$. It is thus a
non-trivial problem for a given supersymmetric model to identify the
flat directions, and the level $n$ of the non-renormalizable superpotential
terms required to lift each flat direction.

In this paper we will study the (nearly) flat directions of the MSSM.
The obvious relevance of understanding the structure of the scalar potential
of this model (which may well describe nature up to a scale
$M_U \sim 2 \times 10^{16}$ GeV at which the gauge couplings appear
[\cite{unification}] to unify)
has recently been highlighted by Dine, Randall and Thomas [\cite{DRT}] in
the context of baryogenesis[\cite{AD}]. Here
we will provide a complete catalog of MSSM flat directions, and
the level at which they are expected to be lifted by non-renormalizable
terms in the superpotential. We will assume that all non-renormalizable
couplings are ``generic'', i.e.,~not subject to any constraints other
than gauge invariance and $R$-parity.
The rest of this paper is organized as follows. In section 2 we will
describe our
notation for the MSSM and provide a basis of gauge-invariant monomials
which parameterize all $D$-flat directions.
These monomials are subject to redundancy relations which are easily
understood in terms of identities obtained by antisymmetrizing over
$SU(3)_C$ and $SU(2)_L$ indices. We will then identify a smaller
basis of monomials which parameterize all renormalizable $F$-flat and
$D$-flat directions. In section 3 we will study how each of the renormalizable
flat directions associated with the monomials identified in section 2
are lifted by non-renormalizable terms in the superpotential.
Most of the flat directions are lifted already at the $n=4$ level by
non-renormalizable terms in the superpotential. However,
we will show that there exists a unique flat direction (which carries
$B-L = 1$) which is not lifted by non-renormalizable operators
until the $n=9$ level, and one other  flat direction (which carries $B-L=-3$)
which is not lifted until the $n=7$ level.
We will also identify other flat directions
which survive until the $n=5$ and $n=6$ levels. Section 4 contains
some concluding remarks.

\subhead{2. Renormalizable flat directions of the MSSM}
\taghead{2.}

Let us begin by specifying our notation and assumptions regarding
the MSSM. The chiral superfields consist of three families of
$SU(2)_L$-doublet quarks $Q$ and leptons $L$ and $SU(2)_L$-singlet
quarks and leptons $u,d,e$, and two Higgs superfields $H_u$ and $H_d$
which couple respectively to up- and down- type quark superfields.
We will use the same symbol for chiral superfields and for their
scalar components.
We will often be able to suppress gauge and family indices
in the following, but when necessary, Greek letters
$\alpha,\beta,\gamma, \ldots$
will be used to refer to $SU(2)_L$ indices; Latin letters $a,b,c \ldots$
to refer to $SU(3)_C$ indices; and Latin letters $i,j,k \ldots = 1,2,3$
to refer to family indices.
All interactions (including soft and non-renormalizable ones)
are assumed to be invariant under an exact $Z_2$
matter parity which is trivially related to $R$-parity by a minus sign for
fermions and which is defined by $(-1)^{3(B-L)}$.
[This assumption follows most naturally [\cite{automatic}] in models
in which $B-L$ is gauged at very high energies and is broken by order
parameter(s) with only even values of $3(B-L)$.]
The renormalizable superpotential is given by
$$
W_{\rm renorm}
= \mu H_u H_d + y_u^{ij} H_u Q_i u_j + y_d^{ij} H_d Q_i d_j +
y_e^{ij} H_d L_i e_j \> .
\eqno(MSSMsupe)
$$
We will assume in the following that the $3\times 3$ Yukawa matrices
$ y_u^{ij},  y_d^{ij}, y_e^{ij}$
are each non-degenerate (have rank 3), although it is worth noting that
this assumption is perhaps not inevitable.

The configuration space of the scalar fields of the MSSM has 49 complex
dimensions (18 for $Q_i$; 9 each for $u_i$ and $d_i$; 6 for $L_i$;
3 for $e_i$; and 2 each for $H_u$ and $H_d$).
The subspace of $D$-flat directions on which the 12 real $D$-term
constraints [8 for $SU(3)_C$; 3 for $SU(2)_L$; and 1 for $U(1)_Y$]
are satisfied is therefore 37 complex dimensional [\cite{DRT}], after 12
corresponding phase degrees of freedom are gauge-fixed. We wish to find
a basis $B$ of gauge-invariant monomials with the property that
any gauge-invariant polynomial in $(Q,$$L,$$u,$$d,$$e,$$H_u$$,H_d)$
can be written as a polynomial in the elements of $B$.
The number of distinct monomials in $B$ is necessarily much greater than
37, because they will be subject to many non-linear redundancy constraints.
The space of $D$-flat directions of the MSSM will then be parameterized
by the elements of the basis $B$, subject to this finite
set of constraints. Our strategy for constructing $B$ is as follows.
First we will construct a basis $B_3$ of
$SU(3)_C$-singlet monomials which transform under $SU(2)_L$ as
singlets, doublets, and a single spin-$3/2$ representation.
Using the elements of $B_3$ as building blocks, we can then construct
a basis $B_{32}$ of monomials which generate all $SU(3)_C \times SU(2)_L$
invariant polynomials. Finally we can combine the elements of $B_{32}$
into weak-hypercharge singlets
to find the basis $B$ of monomials invariant under the full gauge group.

Under $SU(3)_C$, the chiral superfields of the MSSM transform as
13 singlets $(e_i,$ $ L_i,$ $ H_u,$ $ H_d$$)$, six $\bf 3$'s $(Q_i)$, and
six   ${\bf \overline 3}$'s $(u_i, d_i)$. It is useful to adopt
temporarily a generic notation $q_{\sI}$ for ${\bf 3}$'s and $\overline
q_{\sI}$
for ${\bf \overline 3}$'s of $SU(3)_C$, with $I = 1 \ldots 6$.
Then all $SU(3)_C$-invariant polynomials in the $q_{\sI}$ and $\overline
q_{\sI}$
are generated by the 76 monomials
$$
\eqalign{
(q_{\sI} {\overline q}_{\sJ}) & \equiv q_{\sI}^a {\overline q}_{a{\sJ}}
\cr
(q_{\sI} q_{\sJ} q_{\sK}) & \equiv q_{\sI}^a q^b_{{\sJ}} q^c_{\sK}
\epsilon_{abc}
\cr
({\overline q}_{\sI} {\overline q}_{\sJ} {\overline q}_{\sK}) &
\equiv {\overline q}_{a{\sI}} {\overline q}_{b{\sJ}} {\overline q}_{c{\sK}}
\epsilon^{abc}
\> .\cr }
$$
These monomials are not all independent, but are subject to the constraints
following from
$$
q_{\sI}^a (q_{\sK} q_{\sL} q_{\sM})  =
q_{\sK}^a (q_{\sI} q_{\sL} q_{\sM}) +
q_{\sL}^a (q_{\sK} q_{\sI} q_{\sM}) +
q_{\sM}^a (q_{\sK} q_{\sL} q_{\sI})
\eqno(su3constraint1)
$$
and the analogous relation for $\overline q$'s, and by
$$
(q_{\sI} q_{\sJ} q_{\sK}) ({\overline q}_{\sL} {\overline q}_{\sM}
{\overline q}_{\sN}) =
(q_{\sI} {\overline q}_{\sL}) (q_{\sJ} {\overline q}_{\sM} )
(q_{\sK} {\overline q}_{\sN}) \pm
({\rm permutations})
\eqno(su3constraint2)
$$
In \(su3constraint1), the free color index $a$ may be contracted with those
from additional superfields to form either
a $(q{\overline q})$ or a $(qqq)$.
The identity \(su3constraint2) shows that any $SU(3)_C$-invariant polynomial
can be written in a form in which no term contains both a
$(qqq)$ and a $({\overline q}{ \overline q} {\overline q})$.

Now, a color-singlet product of
three $Q$'s can transform under $SU(2)_L$ as either a $\bf 2$ or
a $\bf 4$ of $SU(2)_L$, since
${\bf 2} \times {\bf 2}\times{\bf 2} = {\bf 2} + {\bf 2} + {\bf 4}_S$.
If (QQQ) transforms as a $\bf 4$ of $SU(2)_L$, it must be totally
antisymmetric on its family indices, since it is antisymmetric on
color indices and symmetric on $SU(2)_L$ indices. There is therefore
a unique $SU(3)_C$-singlet monomial made out of three $Q$'s which is a
$\bf 4$ of $SU(2)_L$, namely
$$
(QQQ)^{(\alpha\beta\gamma )}_4 \equiv
Q_i^{\alpha a} Q_j^{\beta b} Q_k^{\gamma c}
\epsilon_{abc} \epsilon^{ijk}
\> .
\eqno(QQQ4s)
$$
The remaining combinations of $(QQQ)$ which are $SU(3)_C$-invariant
are $SU(2)_L$ doublets and can be written in the form
$$
(Q_iQ_jQ_k)^\alpha
\equiv Q_i^{\beta a} Q_j^{\gamma b} Q_k^{\alpha c} \epsilon_{abc}
\epsilon_{\beta\gamma}
\eqno(QQQ2s)
$$
subject to the constraints that not all three of the
family indices may be the same. The basis $B_3$ need not include
terms of the form \(QQQ2s) with $i=k$ or $j=k$, since it is easy to show
that such monomials can be written as linear combinations of terms of
the same type with $i=j$. In fact, there are only 8 linearly independent
$SU(2)_L$-doublet $(QQQ)^\alpha$ monomials, since $(Q_i Q_j Q_k)^\alpha$
is symmetric under
interchange of $i$ and $j$, and $(Q_3 Q_2 Q_1)^\alpha$ is
a linear combination
of $(Q_1 Q_2 Q_3)^\alpha$ and $(Q_1 Q_3 Q_2)^\alpha$. The family indices on
monomials of the form $(uuu)$, $(uud)$,
etc. are subject to the obvious antisymmetrization constraints that
follow from their definitions.

We therefore find that a basis $B_3$ of monomials which generate all
$SU(3)_C$-singlet polynomials of chiral superfields is as given in
Table 1. The gauge index structure is now unambiguous and so is suppressed.
The family indices are suppressed, but take on values 1,2,3
on each chiral superfield (in any convenient basis),
except in the case of $(QQQ)_4$ and with
other restrictions as just discussed.
The $SU(2)_L \times U(1)_Y$ quantum numbers for the monomials and the number
of linearly independent monomials of each type (taking family structure
into account) are also listed in Table 1.
Note that the basis of $SU(3)_C$-invariant monomials contains
23 $\bf 1$'s, 31 $\bf 2$'s and one $\bf 4$ of $SU(2)_L$.

We can now construct a monomial basis $B_{32}$ for the
$SU(3)_C\times SU(2)_L$-singlet polynomials of chiral superfields,
by combining the elements of Table 1 into $SU(2)_L$-singlet
combinations. This is made much easier with the realization that
any term in which an $SU(2)_L$ index from a $(QQQ)_4$
is contracted with another $Q$ can always be rewritten as a polynomial
in terms
of $SU(2)_L$-doublet $(QQQ)$'s instead. This fact can be proved directly by 
examining each possible such term
that can arise; it is interesting to note that if there were 4 or more
families, explicit calculation shows that some terms of this form cannot be 
rewritten in the appropriate way. Therefore,
our monomial basis only includes terms in which the $SU(2)_L$ indices of
$(QQQ)_4$ are contracted with color singlet chiral superfields
$H_u$, $H_d$, or $L$.
Besides such terms, the $SU(2)_L$-singlet monomials include terms
of the form
$$
(\varphi_{\sI} \varphi_{\sJ}) \equiv \varphi_{\sI}^\alpha \varphi_{\sJ}^\beta
\epsilon_{\alpha\beta}
$$
where the $\varphi_{\sI}$ are any of the $SU(2)_L$ doublets in Table 1.
These monomials are clearly antisymmetric under $I \leftrightarrow J$ and
subject to the constraint relations
$$
(\varphi_{\sI} \varphi_{\sJ}) (\varphi_{\sK} \varphi_{\sL}) =
(\varphi_{\sI} \varphi_{\sK}) (\varphi_{\sJ} \varphi_{\sL}) +
(\varphi_{\sI} \varphi_{\sL}) (\varphi_{\sK} \varphi_{\sJ})
\> .
\eqno(su2constraint)
$$
The complete basis $B_{32}$ of monomials which generate all
$SU(3)_C \times SU(2)_L$-singlet polynomials is given in Table 2,
along with their $U(1)_Y$ quantum numbers, which take on the
values $-2,-1,0,1,2$. In Table 2 we have adopted the following
notational conventions. Each consecutive triplet $QQQ$ reading
left-to-right in each monomial is assumed to form a color singlet,
as are consecutive pairs $Qu$ and $Qd$ and consecutive triplets
$uuu$, $uud$, etc.
If a $QQQ$ is not enclosed with parentheses
and a subscript $4$, it is assumed to form a $\bf 2$
of $SU(2)_L$, with the $SU(2)_L$ indices contracted as in \(QQQ2s).
The contractions of the remaining
$SU(2)_L$ indices within each monomial in Table 2 are then uniquely determined.
[For example, in $(QQQ)_4LLL$, the three $L$'s must form a $\bf 4$ of
$SU(2)_L$ in order for the monomial to be an $SU(2)_L$ singlet.]
The suppressed family indices may take on values $1,2,3$ (in any convenient
family basis)
with constraints as discussed before. In many cases, the number of
linearly independent monomials in the basis can be reduced by using
identities \(su3constraint1),
\(su3constraint2), and \(su2constraint). The remaining
basis elements are subject to further non-linear constraints following
from the same identities. In the case of monomials involving $Q$,
one often must use these identities several times in order to
obtain a non-linear redundancy constraint which is written explicitly
in terms of monomials in the basis $B_{32}$.
These redundancy constraints are again themselves highly redundant.

Now we are ready to find the basis $B$ of
$SU(3)_C \times SU(2)_L \times U(1)_Y$-invariant monomials, by combining
elements of Table 2 into $U(1)_Y$-singlets. Referring to the elements
of $B_{32}$ in the generic form $\chi_{-2},\chi_{-1},\chi_0,\chi_1,\chi_2$
with the subscript indicating the weak hypercharge, it is clear that
$B$ consists of monomials of the form $\chi_0$; $\chi_1 \chi_{-1}$;
$\chi_2 \chi_{-2}$; $\chi_2 \chi_{-1} \chi_{-1}$; and
$\chi_{-2} \chi_1 \chi_1$.
(Note that there is exactly one $\chi_2$ and one $\chi_{-2}$.)
However, using the relations \(su3constraint1),
\(su3constraint2), and \(su2constraint), one can show that many of the terms
formed in this way are actually polynomials of other monomials in the
basis. After eliminating as many redundancies as possible in this way,
we find the list of monomials in $B$ given in Table 3.
The conventions for contracting the suppressed gauge indices are
as discussed above for Table 2.
This basis is quite overcomplete, in the sense that there are many
non-linear constraints relating the basis elements following from
\(su3constraint1), \(su3constraint2), and \(su2constraint),
as well as
$$
\eqalign{
(\chi_1^{\sI} \chi_{-1}^{\sJ}) (\chi_1^{\sK} \chi_{-1}^{\sL} ) &=
(\chi_1^{\sI} \chi_{-1}^{\sL}) (\chi_1^{\sK} \chi_{-1}^{\sJ} ) \> ;\cr
(\chi_{-2} \chi_1^{\sI} \chi_{1}^{\sJ})
(\chi_{-2} \chi_1^{\sK} \chi_{1}^{\sL} ) &=
(\chi_{-2} \chi_1^{\sI} \chi_{1}^{\sL})
(\chi_{-2} \chi_1^{\sK} \chi_{1}^{\sJ} ) \> ;\cr
&\vdots }
\eqno(u1constraints)
$$
etc. It is a trivial, if tedious, exercise to write out
all of these redundancy constraints explicitly in terms of the monomials
appearing in Table 3.
It should be kept in mind that these redundancy constraints are themselves
highly redundant.

Each polynomial formed out of the elements of Table 3 corresponds to
a $D$-flat direction of the MSSM.
Though the correspondence between flat directions and gauge-invariant
operators has been discussed extensively in the literature in
simpler models, a short example may be in order. Consider the flat
directions associated with the leptonic sector of the MSSM alone.
In the absence of a superpotential or soft breaking terms, the
scalar potential in this sector is:
$$
\eqalign{
V &= {g^2\over 2}\left ( D_1^2 + D_2^2 + D_3^2 \right ) +
{g^{\prime 2} \over 2 } D_Y^2
\cr
D_1 & = {1\over 2} \sum_i
( L_i^{\uparrow *} L_i^\downarrow + L_i^{\downarrow *} L_i^\uparrow );
\qquad\>\>\>\>\>  D_2 = {i\over 2} \sum_i
( L_i^{\uparrow *} L_i^\downarrow - L_i^{\downarrow *} L_i^\uparrow );
\cr D_3 &= {1\over 2} \sum_i
( |L_i^\uparrow |^2 - |L_i^\downarrow |^2 );
\qquad\>\>\> D_Y = {1\over 2} \sum_i
( 2 |e_i|^2 - |L_i^\uparrow |^2 - |L_i^\downarrow |^2 ).
\cr
}
$$
It is easy to see that $D_Y=D_1=D_2=D_3=0$ and thus
$V=0$ for a class of flat directions of the form
$$
L_i = \left ( {\phi \atop 0} \right ); \qquad
L_j = \left ( {0 \atop \phi} \right ); \qquad
e_k = \phi
$$
with $i\not= j$,
where $\phi$ is the sliding VEV along the flat direction.
Each such flat direction is labeled by a gauge-invariant monomial
$L_i L_j e_k$. This notation is useful for two reasons. First,
the correspondence between $D$-flat directions and gauge-invariant monomials
conveniently obviates the necessity of directly
solving the non-linear equations \(Dflat). Secondly, the $F$-flatness
conditions \(Fflat) can be directly imposed as constraints on
the gauge-invariant operators.

The renormalizable flat directions of the MSSM correspond to the
gauge invariant monomials in Table
3, subject to the additional constraints
$$
\eqalignno{
F^\alpha_{H_u} &= \mu H_d^\alpha + y_u^{ij} Q^\alpha_i u_j = 0 &(fhu)\cr
F^\alpha_{H_d} &= -\mu H_u^\alpha +
y_d^{ij} Q^\alpha_i d_j + y_e^{ij} L^\alpha_i e_j = 0
&(fhd) \cr
F_{Q_i}^{a \alpha} &=
y_u^{ij} H^\alpha_u u^a_j + y_d^{ij} H_d^\alpha d^a_j = 0 &(fQ) \cr
F_{L_i}^{\alpha} &= y_e^{ij} H_d^\alpha e_j = 0 &(fL) \cr
F^a_{u_i} &= y_u^{ji} H_u Q^a_j = 0 &(fu) \cr
F^a_{d_i} &= y_d^{ji} H_d Q^a_j =0 &(fd) \cr
F_{e_i} &= y_e^{ji} H_d L_j = 0 &(fe) \cr
}
$$
Using the assumed invertibility of the Yukawa matrices $y_u$, $y_d$ and
$y_e$, one finds from the $F_u=0$, $F_d=0$ and $F_e=0$ constraints that
any monomial containing an $SU(2)_L$ contraction of $H_d$ with $L$ or $Q$,
or of $H_u$ with $Q$, is immediately constrained to vanish. (A flat direction
is lifted when the corresponding gauge-invariant operator is constrained
to vanish.)
The $F_L=0$ constraint shows that an $H_d$ and $e$ cannot coexist in
any flat direction, while $F_Q=0$ yields a more complicated
constraint on terms containing $uH_u $ and $dH_d $.
Note that at the renormalizable level,
only the four complex constraints following from
$F^\alpha_{H_u}=0$ and $F^\alpha_{H_d}=0$ can lift flat directions
which do not contain a Higgs field.

In the last column of Table 3, we have indicated with a check mark
the monomial flat directions which are always lifted by renormalizable
$F$-term constraints [namely \(fL)-\(fe)]
regardless of family index structure. The only remaining monomials in the
basis which involve the Higgs fields are $L H_u$ and $H_u H_d$.
The moduli space of
renormalizable flat directions of the MSSM is now parameterized
by the gauge-invariant monomials in Table 3
without check marks, subject to the redundancy constraints implied
by \(su3constraint1), \(su3constraint2), \(su2constraint),
\(u1constraints), and the additional
constraints obtained by contracting \(fhu)-\(fe) with
additional chiral superfields to form gauge
singlets. For most applications, including cosmological ones,
one may neglect the contribution to \(fhu) and \(fhd) from the
$\mu$ term, since for phenomenological reasons $\mu$ should
be of order the electroweak scale
$m_W$ and thus its contributions to the scalar potential
at large field strength $|\phi | \gg m_W$ are suppressed by
$m_W/|\phi |$ compared to those from the dimensionless
couplings.

\subhead{3. Lifting of the flat directions by the %
non-renormalizable superpotential}
\taghead{3.}

The MSSM is presumably an effective theory, valid only at scales below
the physical cutoff $M$ which might be associated with a GUT
($M\sim 10^{16}$ GeV) or perhaps a supergravity or superstring model
($M\sim 10^{19}$ GeV). In any case, the new physics associated with the
scale $M$ will give rise to non-renormalizable terms suppressed by
powers of $M$, as indicated schematically by \(power).
Throughout this section, we will assume that all terms consistent with
the gauge symmetries of the MSSM and matter parity will be generated
by short distance effects, with generic couplings of order 1. This is
tantamount to rejecting the possibility of (nearly) exact
global symmetries (including continuous $R$-symmetries) or additional gauged
symmetries left unbroken below the scale $M$.

The problem of characterizing all of the non-renormalizable operators
which can appear in the superpotential is nearly equivalent to the
problem of finding $D$-flat directions, solved in section 2. All
such non-renormalizable operators can be generated using the monomials
appearing in Table 3, with the additional proviso due to matter
parity conservation that each superpotential term contains an
even number of odd matter parity fields $(Q,L,u,d,e)$. In this section
we will consider how the renormalizable flat directions are lifted by
such non-renormalizable superpotential interactions. Since the validity
of the expansion \(power) presumes that the field strengths are less than
$M$, it is sufficient at least formally
to consider the various $F=0$ constraints separately
level-by-level in $n$. At each level $n$, the surviving moduli
space of flat directions is described by the basis of monomials as before,
subject to additional constraints obtained by contracting $F=0$
with additional fields to form polynomials in the basis monomials.
Some of the additional constraints obtained in this way will be
linear in the basis monomials, but in general one finds that
some of the $F=0$ constraints can only be realized non-linearly
on the basis monomials.

It is convenient to start by considering the flat directions
associated with the monomials $H_u H_d$ and $L H_u$ which
contain Higgs fields. (All flat directions involving Higgs
fields are lifted already considering only the renormalizable $\mu$ term,
but as we remarked at the end of the previous section
it is appropriate in many applications to
neglect $\mu \approx m_W$.) The $n=4$ superpotential includes terms
$$
W_4 \supset
{\lambda\over M} (H_u H_d)^2 +
{\lambda^{ij}\over M} (L_i H_u) (L_j H_u)
\eqno(w4higgs)
$$
The $F_{H_d}=0$ constraint following from \(w4higgs)
is
$\lambda H_u^\alpha (H_u H_d) = 0$, which upon contraction
with $\epsilon_{\alpha\beta} H_d^\beta$ immediately implies
$H_u H_d = 0$ if $\lambda \not= 0$. The constraint
$F_{H_u} = 0$ then similarly implies, after multiplying by $H_u$, that
$\lambda^{ij} (L_i H_u)(L_j H_u) = 0$, which in turn requires
$L_i H_u=0$ for all $i$ as long as det$\lambda^{ij} \not= 0$.
Since the terms in \(w4higgs) are the only ones allowed in the
$n=4$ superpotential which involve $H_u$ and $H_d$, it is clear
that there can be no interference between the $F$-constraints on
the monomials $H_u H_d$ and $L_i H_u$ and those on the remaining
monomials in the renormalizable flat basis. Thus we find that
with generic couplings $\lambda \not=0$ and det$\lambda^{ij} \not= 0$,
all flat directions involving $H_u$ or $H_d$ are lifted by $n=4$
terms in the superpotential, i.e., dimension 6 terms in the scalar
potential.

The remaining renormalizable flat directions which do not involve
Higgs fields are described by gauge-invariant  monomials
$$
\matrix{
LLe; \qquad & udd; \qquad & QdL; \qquad & QQQL; \qquad & QuQd; \cr
uude; \qquad & QuLe; \qquad & dddLL; \qquad & uuuee; \qquad & QuQue;\cr
QQQQu; \qquad & (QQQ)_4 LLLe; \qquad & uudQdQd \> . \qquad & {} & {}\cr
}
$$
 With all monomials containing Higgs fields already constrained
to vanish, the only superpotential terms which can play a role in
lifting these are ones which contain neither $H_u$ nor $H_d$
(if $n$ is even), or exactly one of $H_u$ or $H_d$ (if $n$ is odd),
because of matter parity conservation.
For example, at $n=4$  the part of the superpotential which is pertinent for
flat directions not involving Higgs fields is just a sum over the
213 linearly independent monomials of the correct dimensionality:
$$
\eqalign{
& W_4  = W_4^{QQQL} + W_4^{QuQd} + W_4^{QuLe} + W_4^{uude}; \cr
& W_4^{QQQL} = \sum_{I=1}^{24} {\alpha_I \over M} (QQQL)_I; \qquad
\>\> W_4^{QuQd} = \sum_{I=1}^{81} {\beta_I \over M} (QuQd)_I; \cr
& W_4^{QuLe} = \sum_{I=1}^{81} {\gamma_I \over M} (QuLe)_I; \qquad
\>\> W_4^{uude} = \sum_{I=1}^{27} {\delta_I \over M} (uude)_I\> . \cr
}\eqno(W4)
$$
For odd $n$, only $F_{H_u}=0$
or $F_{H_d}=0$ constraints can help to lift the remaining flat
directions, while for even $n$, only $F_Q=0$, $F_L=0$,
$F_u=0$, $F_d=0$, and $F_e=0$ impose constraints on the remaining
polynomials.

The existence of flat directions can be viewed as an ``accidental''
consequence of the limitations imposed on the superpotential at
a given level $n$
by the requirements of gauge invariance and matter parity. As discussed
in the Introduction, the number of $F$-constraints always
formally exceeds the
number of gauge-invariant degrees of freedom, so that flat directions
only occur because in certain exceptional
directions in field space, the $F$-constraints are trivially satisfied.
It might seem intuitively clear, then, that the flat directions which have
the greatest possibility of surviving unlifted to large $n$ are those
which involve only a few different superfields, since in such
special directions many of the $F$-terms vanish automatically.
This intuition works well in most cases, but as we shall
see, it suffers one major exception in the MSSM. In the
following discussion, we study first the
cases of flat directions which involve only two different types
of chiral superfields.

\noindent $\bullet$ L,e {\it flat directions}.
Flat directions involving only the chiral superfields $L$ and $e$ are
parameterized by gauge-invariant operators $LLe$.
There are 9 linearly independent monomials of this type,
since there are 3 ways to assign family indices to $LL$ and
3 $e$'s. However, not all of these are functionally independent,
because of the redundancy constraints \(u1constraints), which
in this case take the form
$$
(L_i L_j e_k) (L_{i^\prime} L_{j^\prime} e_{k^\prime}) =
(L_i L_j e_{k^\prime}) (L_{i^\prime} L_{j^\prime} e_{k})
\> .
\eqno(redLLe)
$$
These constraints allow us to solve for the 9 $LLe$ monomials
in terms of just 5 of them, which may be taken to be
$L_1 L_2 e_i$, $L_1 L_3 e_1$, $L_2 L_3 e_1$ in an arbitrary
basis, as long as $L_1 L_2 e_1 \not=0$.
[Alternatively, one may note that there are 6 complex degrees of freedom
in $L$ and 3 in $e$, subject to 4 real gauge constraints from
$SU(2)_L \times U(1)_Y$ $D$-flatness,
leaving 5 complex degrees of freedom after
4 additional phases are gauge-fixed.]
There are 2 independent constraints on these monomials
due to the $F^\alpha_{H_d}=0$ condition following from
the renormalizable superpotential. Therefore the moduli subspace of
renormalizable flat directions of the type $LLe$ has 3 complex dimensions.
None of the $n=4$ superpotential terms can lift the remaining flat directions,
since they all involve at least two fields other than $L$ and $e$, so
that the $F$-terms resulting from them must involve at least one field
other than $L$ and $e$.
At the $n=5$ level the only relevant superpotential terms are of the form
$$
W_5 \supset {1\over M^2}  H_u L L L e \> .
\eqno(W5HLLLe)
$$
Following from this are two independent constraints $F^\alpha_{H_u}=0$ which
after multiplying by $\epsilon_{\alpha\beta}L^\beta e$
are realized non-linearly on the monomials $LLe$. So the moduli
subspace of flat directions which remain unlifted at $n=5$ has
one complex dimension.
Finally at the $n=6$ level the 9 independent constraints $F_L=0$ and $F_e=0$
obtained from
$$
W_6 \supset {1\over M^3} LLe LLe
\eqno(W6LLeLLe)
$$
provide an overconstraining set of requirements on the monomials $LLe$.
Therefore we find that all flat direction of the type $LLe$ are lifted
by the $n=6$ non-renormalizable superpotential.

\noindent $\bullet$ u,d {\it flat directions}.
Flat directions involving only $u$ and $d$ fields are labeled by the
gauge-invariant monomials $udd$.
There are 9 linearly (and functionally) independent such monomials
in the basis $B$.
Clearly, none of the renormalizable ($n=3$) superpotential terms can
yield $F$-constraints which affect these monomials.
At the $n=4$ level, the relevant part of the superpotential is
$W_4^{uude}$ of \(W4).
The resulting $F_{e}=0$ constraints are realized non-linearly
on the $udd$ monomials, after
multiplying by $ddd$ and using \(su3constraint1), schematically:
$$
ddd {\partial \over \partial e} W_4^{uude} = \sum (udd) (udd) = 0
\> .
\eqno(heffalump)
$$
Since there are only 3 such constraints that are functionally independent,
(one for each of the $e_i$ with respect to which the derivatives are taken),
the moduli subspace of directions of the $udd$ type which remain
flat at the $n=4$ level has 6 complex dimensions. At the $n=5$ level
of the superpotential, there are no available $F$-terms which
can constrain the monomials $udd$ by themselves. At the $n=6$ level,
one finds terms of the form
$$
W_6 \supset {1\over M^3} uddudd \> .
\eqno(W6uddudd)
$$
The 18 independent constraints $F_u=0$ and $F_d=0$ following from
this will clearly overconstrain the remaining 6 complex degrees of freedom,
so that all monomials $udd$ must vanish. So we find that all flat
directions involving $u,d$ are lifted at the level
$n=6$.

\noindent $\bullet$ Q,L {\it flat directions}.
Flat directions involving only fields $Q$ and $L$ are associated with
gauge-invariant monomials $QQQL$. There are
24 linearly independent monomials of the type $QQQL$.
[Recall that there are 8 linearly independent doublets $QQQ$ and three $L$s,
and that the results of the previous section show that monomials involving
$(QQQ)_4$ can be eliminated except in terms which involve Higgs fields or $e$,
which are not in question here.]
Using the identities \(su3constraint1) and \(su2constraint), one
can show
that only 12 of these monomials are actually functionally independent.
[This can alternatively be understood by the following counting:
there are 18 independent Q's and 6 independent L's, subject to
12 real gauge constraints from $SU(3)_C$, $SU(2)_L$, and $U(1)_Y$,
and 12 phase gauge-fixings.]
None of the $F$-constraints from $n=3$ superpotential terms can
restrict these monomials. At the $n=4$ level, the relevant superpotential
terms are just those which have the same form as the monomials
we are trying to lift, namely $W_4^{QQQL}$ of \(W4).
The 24 functionally independent $F$-constraints
$$
{\partial \over \partial Q_i^{\alpha a}} W_4^{QQQL} = 0;
\qquad
{\partial \over \partial L_i^{\alpha}} W_4^{QQQL} = 0
\eqno(QQQLF)
$$
ought to overconstrain the 12 functionally independent variables
$QQQL$. Let us demonstrate the validity of this sort of counting argument
by proving explicitly that all of the flat directions $QQQL$ are indeed
lifted.
{}From \(QQQLF) one obtains constraints linear in the monomials
$QQQL$:
$$
Q_j {\partial \over \partial Q_i}  W_4^{QQQL} =
\sum (QQQL) =0; \qquad\>
\>
L_j {\partial \over \partial L_i}  W_4^{QQQL} = \sum (QQQL) =0 \> .
\eqno(QQQLF2)
$$
Of these 18 equations, 17 are linearly independent. In practice,
one can choose random numerical values for the 24 couplings
$\alpha_I$, and use \(QQQLF2) to solve for the 24 $QQQL$ monomials
in terms of 7 unknowns $x_1 \ldots x_7$. [The last constraint
encountered in \(QQQLF2) will always amount to $0=0$.]
Then the non-linear redundancy constraints \(su3constraint1)
can be rewritten using \(su2constraint) in the form
$$
\sum (QQQL)(QQQL) = 0
\eqno(QQQLred) $$
which become homogeneous
quadratic equations in the 7 unknowns $x_1\ldots x_7$.

Now, there do exist mathematical algorithms
well-known in algebraic geometry (e.g.~the methods
of resultant polynomials [\cite{resultants}]
or Gr\"obner bases [\cite{grobner}])  which can in
principle be used to decide whether such systems of
simultaneous non-linear equations have
non-trivial solutions. Unfortunately, these methods are of no
practical use at the level of complexity encountered here or in the
examples below; the
required number of algebraic operations is demonstrably finite,
but quite astronomical.

In the present example, fortunately, it is still possible to prove
rigorously and explicitly using a more brutish method
that there is generically no non-trivial solution to the constraints.
One may simply treat the 28 unknowns $z_{ij} \equiv x_i x_j$
as independent variables. Then one can show that 20 of the equations
\(QQQLred) are linearly independent in the $z_{ij}$.
To complete the proof, one may note that we have so far
only used 17 of the 24 functionally independent equations \(QQQLF).
Additional constraints are obtained from
$$
LLQQQ^{\alpha} {\partial \over \partial L^\alpha} W_4^{QQQL}
= \sum (QQQL)(QQQL) = 0
\eqno(tigger)
$$
These constraints are quadratic in the monomials $QQQL$, hence
linear in the $z_{ij}$. One can check that indeed exactly $24-17 =7$
of these constraints are linearly independent of each other and
the previous constraints, in terms of the $z_{ij}$. Thus one can solve
for all of the $z_{ij}$ in terms of, say, $z_{11}$. Finally one can compute
$z_{11} z_{22} - z^2_{12}$, which is constrained to be 0 by construction.
Written in terms of the last remaining unknown $z_{11}$, it is easy
to check that only the trivial solution $z_{11} = 0$ exists.
Hence, all of the $QQQL$ monomials are constrained to be zero,
so that all of the flat directions associated with them are
indeed lifted by the $n=4$ superpotential.

\noindent $\bullet$ d,L {\it flat directions}.
Flat directions involving just the fields $d,L$ are labeled by
the gauge-invariant  monomials $dddLL$.
These flat directions are
a particularly exceptional case, because few terms in the superpotential
involve only $d$ and $L$ and just one other chiral superfield.
(Superpotential terms containing two or more fields other than $d$ and $L$
will result in $F$-terms which cannot constrain
the $dddLL$ monomials, except in combinations with other monomials.)
The number of linearly independent monomials $dddLL$ is 3, because
there is only one color-singlet combination $ddd$, and 3 possible
family index assignments for $LL$. All three monomials $dddLL$ are also
functionally independent, in the sense that none of them can be
eliminated by using the non-linear redundancy identities.
It is easy to see that the lowest dimension term in the superpotential
which can lift flat directions of the type $dddLL$ is the $n=5$ term
$$
W_5 \supset {\lambda^i \over M^2} ddd H_d L_i
\eqno(W5dddHL)
$$
The requirement $F^\alpha_{H_d}=0$ following from this is just
$$
\lambda^i ddd L_i^\alpha = 0
\eqno(dddLLcon)
$$
which yields only 2 independent constraints on the 3 monomials $dddLL$.
Therefore, if $\lambda^i \not= 0$, exactly one of the
flat directions of the type $dddLL$ is not lifted at the level $n=5$.
To make this transparent, one may choose a basis in family space such that
$\lambda^i \propto \delta^{i1}$.
Then clearly  \(dddLLcon) forces the monomials $dddL_1L_2$ and
$dddL_1 L_3$ to vanish, but can never constrain the monomial
$dddL_2L_3$. There is no possibility of
$F$-constraints on $dddLL$ flat directions
coming from $n=6$ terms in the superpotential, since these all
involve at least two fields other than $d$ and $L$, so that at least
one field other than $d$ and $L$ remains after taking the derivative.
At the $n=7$ level, one finds superpotential terms of the form
$$
W_7 \supset {1\over M^4} H_u LLL ddd
\eqno(W7)
$$
The two independent $F^\alpha_{H_u}=0$ constraints following from this term
are realized non-linearly on the monomials $dddLL$ (after multiplying
by $dddL$) and finally
lift the last remaining $dddLL$ flat direction.
Therefore we find that exactly one of the $dddLL$ flat directions
is not lifted until the $n=7$ level.

\noindent $\bullet$ u,e {\it flat directions}.
The flat directions involving only the fields $u$ and $e$ correspond to
gauge-invariant monomials $uuuee$.
There are 6 independent monomials of the type $uuuee$, since
there is only one color-singlet combination $uuu$, and 6 independent
products $e_i e_j$. Of these, only three combinations
are functionally independent, because of identities of the type
\(u1constraints), e.g.,~$(uuue_1 e_2)^2 = (uuue_1e_1)(uuue_2e_2)$.
The only superpotential terms with $n<9$ which can yield
$F$-terms constraining these monomials alone is
$W_4^{uude}$ of \(W4). It is not difficult to show that
the 9 independent constraints
$$
{\partial \over \partial d^a_i} W_4^{uude} = 0
\eqno(uuueecon)
$$
are sufficient to require all $uuuee$ monomials to be zero.
[One can simply  consider the linear equations obtained
by contracting \(uuueecon) with $u^a_j e_k$, with
random numerical values for the couplings
$\delta_I$ in \(W4).] Therefore we find that
all flat directions associated with the monomials $uuuee$ are
lifted by the non-renormalizable superpotential with $n=4$.

\noindent $\bullet$ Q,u {\it flat directions}.
These are parameterized by the monomials $QQQQu$ in the basis $B$ of
Table 3. One can show, using \(su3constraint1) and \(su2constraint)
that there are 54 linearly independent monomials $QQQQu$. Of these,
15 are functionally independent in the sense that they cannot be
eliminated using non-linear
constraints also following from \(su3constraint1) and \(su2constraint).
At the renormalizable ($n=3$)
level, there are 2 functionally independent $F_{H_u}^\alpha$ constraints,
so that the moduli subspace of renormalizable flat directions involving
$Q$ and $u$ has 13 complex dimensions. At the $n=4$ level, the pertinent
terms in the superpotential are $W_4^{QuQd}$ and $W_4^{QQQL}$ in \(W4).
We obtain 9 functionally independent constraints on the monomials
$QQQQu$ from $F_d$ of $W_4^{QuQd}$, and 6 more from $F_L$ of
$QQQL$. Thus there are a total of 17 independent constraints from
the $n=3$ and $n=4$ superpotentials, which should therefore overconstrain
the $15$ functionally independent degrees of freedom, so that
all of the flat directions are lifted at $n=4$.

This concludes the discussion of the flat directions which involve only
two different types of chiral superfields. Note that there can be no flat
directions of the type $Q,d$ or $Q,e$ or
$d,e$ or $L,u$, since these clearly cannot satisfy the $D_Y=0$
constraint. Next we consider
the flat directions which involve three different types
of chiral superfields.

\noindent $\bullet$ L,d,e {\it flat directions}.
The flat directions involving only $L$, $d$, and $e$ chiral superfields
are associated with the
12 linearly independent monomials $LLe$ and $dddLL$.
Of these, only 6 are functionally independent when the non-linear
constraints \(u1constraints) are taken into account. Of course, these
flat directions contain as two exceptional subspaces the flat directions
$L,e$ (with $ddd=0$) and $d,L$ (with all $e_i=0$),
which have already been discussed, so here we can assume that
$ddd \not=0$ and some $e_i \not=0$ (and of course some $L_iL_j \not=0$),
so that these can be divided by in reducing the number
of independent complex degrees of freedom to 6.
Now at the level of the renormalizable ($n=3$) superpotential,
there are 2 independent $F_{H_d}^\alpha = 0$ constraints.
None of the $n=4$ superpotential terms produces a relevant $F$-constraint.
At the level $n=5$, one finds 2 constraints $F_{H_u}^\alpha = 0$ from
the superpotential term \(W5HLLLe) and 2 independent
constraints $F_{H_d}^\alpha = 0$ from the
superpotential term \(W5dddHL). Thus other than the degenerate
cases discussed above in which either $ddd$ or all $e_i$ vanish,
the $L,d,e$ flat directions are lifted at the
$n=5$ level.

\noindent $\bullet$ L,d,u {\it flat directions}.
These flat directions are labeled by the 12 linearly (and functionally)
independent monomials $udd$ and $LLddd$.
None of the renormalizable superpotential terms help to lift these flat
directions. At the $n=4$ level, one obtains 3 constraints $F_e=0$ from
the superpotential term $W_4^{uude}$ of \(W4). At the $n=5$ level, one finds
2 constraints $F_{H_u}=0$ from the superpotential term
$$
W_5 \supset {1\over M^2} udd H_u L
\eqno(W5uddHL)
$$
and 2 constraints $F_{H_d}=0$ from \(W5dddHL).
At the $n=6$ level one obtains 18 constraints $F_Q=0$ from the superpotential
terms
$$
W_6 \supset {1\over M^3} uddQdL
\eqno(W6uddQdL)
$$
so that the system is overconstrained. Therefore, other than the
exceptional case discussed above with all $u_i=0$,
the $L,d,u$ flat directions are lifted at $n=6$.

\noindent $\bullet$ L,u,e {\it flat directions}.
The flat directions involving $L,u,e$ are parameterized by the 15 linearly
independent monomials $LLe$ and $uuuee$, subject to non-linear constraints
of the type \(u1constraints)
which reduce the number of functionally independent monomials to 6.
At the renormalizable level one has only 2 constraints on
these 6 degrees of freedom from $F_{H_d}=0$.
At the $n=4$ level, the 18 $F_Q=0$ constraints from
$W_4^{QuLe}$
and the 9 $F_d=0$ constraints from $W_4^{uude}$ clearly overconstrain the
system, lifting all of these flat directions at $n=4$, except the
ones treated above with $u_i=0$.

\noindent $\bullet$ u,d,e {\it flat directions}.
These are associated with the 42 linearly independent monomials
$udd$, $uuuee$ and $uude$. Of these, only 12 are functionally independent
after taking into account the non-linear identities of the form
\(su3constraint1) and \(u1constraints), besides the exceptional cases
discussed above in which all $e_i$ vanish or all $d_i$ vanish.
The renormalizable superpotential does not lift any of these terms.
At the $n=4$ level, the 21 independent constraints $F_u=0$, $F_d=0$
and $F_e=0$ following from the superpotential term
$W_4^{uude}$ are enough to lift all flat directions of this type, other than
in the special case treated above that all $e_i$ vanish.

\noindent $\bullet$ Q,u,e {\it flat directions}.
This is a quite exceptional case, for a somewhat subtle reason.
These flat directions are associated with gauge-invariant
monomials $uuuee$, $QQQQu$, and $QuQue$. Using the non-linear
redundancy constraints, one finds that 18 of these monomials are
functionally independent. Now, at the renormalizable level,
there are 2 independent constraints $F_{H_u}$, so that the moduli
subspace of renormalizable flat directions involving $Q,u,e$ has
16 complex dimensions. At the $n=4$ level, one has 6 independent
constraints $F_L$ and 9 independent constraints $F_d$,
which schematically take the form
$$
\eqalignno{
0 = F_L & = \sum (QQQ) + \sum (Que); &(FLque)\cr
0 = F_d & = \sum (QuQ) + \sum (uue) &(Fdque) \cr
}
$$
respectively. These constraints are realized linearly on the monomials
after multiplying \(FLque) by $Qu$ and
\(Fdque) by $QQ$ and $ue$, but only 15 such constraints are
functionally independent.
Therefore, one complex degree of freedom must remain unlifted at the
$n=4$ level!  One can easily check that no superpotential terms with
$5 \le n \le 8$
can affect the $Q,u,e$ flat directions, since they all contain at least
two superfields other than $Q,u,e$. At the $n=9$ level there are many
terms which can lift the last remaining flat direction, for example
$$
W_9 \supset {1\over M^6} QuQuQuH_dee \> .
\eqno(eeyore)
$$
The $F_{H_d}$ constraints coming from this term will clearly lift
the last remaining $Q,u,e$ flat direction.
The unique $Q,u,e$ flat direction which survives to $n=9$
is at least formally
the ``flattest'' of the flat directions of the MSSM, being lifted
only by terms of dimension 16 in the supersymmetry-preserving
part of the scalar potential.
This result is perhaps somewhat surprising, as we found above that
all of the $Q,u$ flat directions and $u,e$ flat directions, which
are special cases of $Q,u,e$ flat directions, are lifted already
at $n=4$. The subtlety is that at the $n=4$ level, the $F$-terms
which were responsible for lifting the $Q,u$ and $u,e$ flat directions are
the same, and cannot be counted separately in the more general case
$Q,u,e$. The specific identification of the flat direction which survives
to $n=9$ depends in an extremely complicated non-linear way on the unknown
$n=4$ couplings.

\noindent $\bullet$ Q,L,d {\it flat directions}.
These flat directions are associated with the 54 linearly independent
monomials $QQQL$, $dddLL$, and $QdL$. Only 21 of these monomials are
functionally independent after taking into account the non-linear
redundancy constraints. At the renormalizable level,
there are just 2 constraints $F_{H_d}=0$ on these 21 degrees of freedom.
At the $n=4$ level, there are 24 constraints $F_Q=0$ and $F_L=0$
from $W_4^{QQQL}$ of \(W4) and 9 more constraints $F_u=0$ from $W_4^{QuQd}$.
Hence all $Q,L,d$ flat directions are lifted at $n=4$, other than the
exceptional case of $L,d$ flat directions as discussed above.

\noindent $\bullet$ Q,u,d {\it flat directions}.
Here the monomials which label the $D$-flat directions are
$udd$, $QQQQu$, $QuQd$, and $uudQdQd$. Of these, 24 can be shown to be
functionally independent using the non-linear redundancy constraints.
At the renormalizable level, there are 4 independent constraints
from $F_{H_u}=0$ and $F_{H_d}=0$. At the $n=4$ level, one has 36
constraints $F_Q=0$, $F_u=0$ and $F_d=0$ from the superpotential
term $W_4^{QuQd}$ of \(W4) and 3 more constraints $F_e=0$ from $W_4^{uude}$.
Thus all $Q,u,d$ flat directions are lifted at $n=4$
except in the case treated above in which the monomials
dependent on $Q$ vanish.

\noindent $\bullet$ Q,L,e; Q,L,u; Q,L,u,e; Q,L,d,e; Q,L,u,d;
Q,u,d,e; L,u,d,e; and Q,L,u,d,e {\it flat directions}.
Using methods which should be clear by now, one can show
that each of these flat directions is lifted by the $n=4$ superpotential,
except in degenerate cases already covered in the above discussion.
Note that there can be no flat directions of the type
$Q,d,e$, since these clearly cannot satisfy the $D_Y=0$
constraint.

Some of the essential features of the preceding discussion are
summarized in Tables 4 and 5. In Table 4, for each type of flat direction
we put a check mark for each superpotential term which
produces a pertinent $F$-constraint. (In ``overkill'' cases where all flat
directions of a given type are already lifted by superpotential terms
with smaller $n$, no entry is made.) In Table 5,
we show the complex dimension of the moduli space associated with
each type of flat direction which remains unlifted at each level $n$.
A check mark indicates the minimum level $n$ at which all flat directions
of a given type are lifted.

\subhead{4. Discussion}
\taghead{4.}

In this paper we have provided a catalog of the (nearly) flat directions
of the MSSM. We found that there exists a unique direction, involving
the fields $(Q,u,e)$ in a non-trivial way, which is formally
``flattest''. It is lifted only by soft supersymmetry-breaking effects
and by $n=9$ terms in the superpotential, which correspond to dimension
16 terms in the scalar potential. The next flattest direction involves
the fields $(L,d)$ and is lifted at the level $n=7$, corresponding to
dimension 12 supersymmetric
terms in the scalar potential. There are several flat directions
with sliding VEVs for fields $(L,e)$, for $(u,d)$, and for $(L,u,d)$
which are lifted at the $n=6$ level corresponding to dimension 10 terms
in the scalar potential, and some flat directions involving $(L,d,e)$
are lifted at the $n=5$ level corresponding to dimension 8 terms in
the scalar potential. All other flat directions are lifted already at
the $n=4$ level, corresponding to dimension 6 terms in the supersymmetric
part of the scalar potential.  Our results seem to
differ slightly with those found in the second reference in [\cite{DRT}],
although this fact is probably quite
peripheral to the main points of that paper.
It should be noted that the notion of ``generic'' couplings may have to
allow for dimensionless numbers which are several orders of magnitude
less than unity; after all, the Yukawa coupling for the electron in
the standard model is less than $10^{-5}$. Therefore the formal distinction
between flat directions lifted at $n=6$ and
at higher $n$ may be quite moot in practice,
in view of our lack of knowledge of the non-renormalizable couplings.
This situation could change in the context of model extensions of the MSSM
which make more specific predictions for the non-renormalizable couplings.

One possible application of our results is to the question of the
existence of unbounded from below directions or non-trivial global
minima of the scalar potential along nearly flat directions.
It is possible that the running (mass)$^2$ parameters of the scalars
of the MSSM could become negative at very high energy scales. Even
within the framework of unified supersymmetric models,
it is conceivable [\cite{KM}]
that the common scalar (mass)$^2$ parameter $m_0^2$
could be negative at the unification scale. This does not necessarily
lead to VEVs for squarks or sleptons. The reason is that
the prediction of such VEVs is only trustworthy if the running
parameters are evaluated at the scale of the would-be VEVs.
The important terms in the scalar potential in this regard are
\(soft) and \(hard). Thus the running (mass)$^2$ parameter
$m^2(Q)$ corresponding to a renormalizable flat direction
lifted at level $n$ must remain positive for, roughly,
$$
Q < \left ( {m_W M^{n-3} \over |\lambda | } \right )^{1/(n-2)}
\eqno(poohbear)
$$
so that equations for a putative VEV can have no solution.
Such a constraint has been examined very recently in [\cite{FORS}]
for one particular flat direction of the type described
by the monomials $udd$.

Terms in the scalar potential of the form
\(softholo) are also potentially dangerous with regard to
unwanted global minima of the scalar potential at large field
strength if $s<n$.
However, it is easy to check that
for the seemingly most-dangerous
$Q,u,e$ and $L,d$ flat directions, one has $s \ge 10$ because
of the requirement of
matter parity invariance. Similarly, for the flat directions
with smaller $n$ one finds that the maximum $n$ never exceeds the
corresponding $s$.

We must emphasize that our results depend on certain controvertible
assumptions about physics at very high energies.  For example, if there
exist at
an intermediate scale below $M$ some chiral superfields with
vector-like quantum numbers, these could
have non-trivial superpotential couplings to the chiral superfields
of the MSSM, lifting
some of the flat directions at smaller $n$ than found here.
As an example, one could imagine introducing gauge-singlet neutrino chiral
superfields $\nu$ so that $W_{\rm renorm}$ would include couplings
schematically of the form $ \mu_{I} \nu\nu + y_{\nu} H_u L \nu$,
implementing a supersymmetric
seesaw mechanism [\cite{seesaw}] for neutrino masses.
At the $n=4$ level, one would then expect to
have matter parity invariant couplings
$$
W_{4} \supset {1\over M} (LLe\nu + udd\nu + QdL\nu ) ,
\eqno(christopherrobin)
$$
and the $F_\nu=0$ constraints
could help to lift some of the flat directions discussed above
which survived beyond $n=4$ in the MSSM alone.
New symmetries can also clearly affect our conclusions. In the case
of global symmetries, some of the couplings assumed
to be generic here could actually
be zero or otherwise restricted, so that flat directions
could remain unlifted to higher $n$.
Conversely, if there are additional unbroken gauged symmetries
above some intermediate scale, the corresponding
$D$-terms would lift many of the flat directions.
Clearly for any extension of the MSSM, the status
of flat directions at very large field strength must be reexamined.
The methods used here may also serve as a helpful guide in such cases.

\noindent Acknowledgments:
We are grateful to Michael Dine, Gordy Kane, John March-Russell
and Lisa Randall for helpful discussions.
The work of TG and SPM was supported in part by the U.S. Department of Energy.
The work of CK was supported in part
by the U.S. Department of Energy under contract \#DE-FG02-90ER40542 and by the
Monell Foundation.

\hyphenation{L-e-i-g-h}
\references

\refis{unification} P.~Langacker, in Proceedings of the
PASCOS90 Symposium, Eds.~P.~Nath
and S.~Reucroft, (World Scientific, Singapore 1990)
J.~Ellis, S.~Kelley, and D.~Nanopoulos, \pl B260, 131, 1991;
U.~Amaldi, W.~de Boer, and H.~Furstenau, \pl B260, 447, 1991;
P.~Langacker and M.~Luo, \pr D44, 817, 1991.

\refis{reviews}
For reviews, see H.~P.~Nilles,  \prpts 110, 1, 1984;
H.~E.~Haber and G.~L.~Kane, \prpts 117, 75, 1985;
H.~E.~Haber, ``Introductory low-energy supersymmetry'', TASI-92 lectures,
hep-ph/9306207.

\refis{automatic} R.N.~Mohapatra, \pr D34, 3457, 1986;
A.~Font, L.E.~Ib\'a\~nez, and F.~Quevedo \pl B228, 79, 1989;
L.~E.~Ib\'a\~nez and G.~G.~Ross, \pl B260, 291, 1991 and
\np B368, 3, 1992; S.~P.~Martin, \pr D46, 2769, 1992.

\refis{grobner} D.~Cox, J.~Little, and D.~O'Shea, ``Ideals, Varieties,
and Algorithms'' Springer-Verlag, New York, (1992).

\refis{resultants} B.~L.~van der Waerden, ``Modern Algebra'', v.~2,
Ungar Publishing, New York, (1950).

\refis{AD} I.~Affleck and M.~Dine, \np B249, 361, 1985.

\refis{ADS} I.~Affleck, M.~Dine, and N.~Seiberg,
\np B241, 493, 1984; {\sl Nucl.~Phys.} {\bf B256}, (1985), 557.

\refis{BDFS} F.~Buccella, J.~P.~Derendinger, S.~Ferrara, and C.~A.~Savoy,
\pl B115, 375, 1982.

\refis{taylor} T.~Taylor, \pl B125, 185, 1983; \pl B128, 403, 1983.

\refis{Seiberg} N.~Seiberg, \pr D49, 6857, 1994;
\np B435, 129, 1995.

\refis{IS} K.~Intriligator and N.~Seiberg, \np B431, 551, 1994;
K.~Intriligator, R.~G.~Leigh, and N.~Seiberg, \pr D50, 1092, 1994.

\refis{PR} E.~Poppitz and L.~Randall, \pl B336, 402, 1994;
J.~Bagger, E.~Poppitz, and L.~Randall, \np B426, 3, 1994.

\refis{seesaw} M.~Gell-Mann, P.~Ramond, and R.~Slansky, in Sanibel Talk,
CALT-68-709, Feb 1979, and in {\it Supergravity},
(North Holland, Amsterdam, 1979; T.~Yanagida, in {\it Proc. of the
Workshop on
Unified Theories and Baryon Number in the Universe}, Tsukuba, Japan, 1979,
edited by A. Sawada and A. Sugamoto (KEK Report No. 79-18, Tsukuba, 1979).

\refis{ILS} K.~Intriligator, R.~G.~Leigh and M.~J.~Strassler,
``New examples of duality in chiral and non-chiral supersymmetric
gauge theories'', preprint hep-th/9506148.

\refis{LT} M.~A.~Luty and W.~Taylor,
``Varieties of vacua in classical supersymmetric gauge theories",
preprint hep-th/9506098.

\refis{GP} S.~B.~Giddings and J.~M.~Pierre, ``Some Exact Results in
Supersymmetric Theories Based on Exceptional Groups'', preprint
hep-th/9506196.

\refis{KM} C.~Kolda and S.~P.~Martin, ``Low-energy supersymmetry
with D-term contributions to scalar masses'', preprint
hep-ph/9503445.

\refis{FORS} T.~Falk, K.~A.~Olive, L.~Roszkowski, and M.~Srednicki,
``New Constraints on Superpartner Masses'', preprint
hep-ph/9510308.

\refis{DRT} M.~Dine, L.~Randall, and S.~Thomas,
\prl 75, 398, 1995; and
``Baryogenesis from Flat Directions of the Supersymmetric Standard Model",
preprint hep-ph/9507453.

\endreferences
\endit\end